\documentclass[epj]{svjour}
\usepackage{graphicx}
\usepackage{amsmath}
\begin{document}
\title{Enhancing energy harvesting by coupling monostable oscillators}
\author{J.I. Pe\~na Rossell\'o\inst{1} \and H.S. Wio\inst{2} \and R.R. Deza\inst{1} \and P. H\"anggi\inst{3,4}}
\offprints{julian@ifimar-conicet.gob.ar}
\institute{Instituto de Investigaciones F\'{\i}sicas de Mar del Plata (IFIMAR, UNMdP\&CONICET), FCEyN-UNMdP,\\De\'an Funes 3350, B7602AYL Mar del Plata, Argentina \and Instituto de F\'{\i}sica de Cantabria (IFCA, UC\&CSIC), Avda.\ de los Castros, s/n, E-39005 Santander, Spain \and Universit\"at Augsburg, Institut f\"ur Physik, Universit\"atstrasse 1, D-86135 Augsburg, Germany \and Nanosystems Initiative Munich, Schellingstrasse 4, 80799 M\"unchen, Germany}
\date{Received: date / Revised version: date}
%
\abstract{The performance of a ring of linearly coupled, monostable nonlinear oscillators is optimized towards its goal of acting as energy harvester---through piezoelectric transduction---of mesoscopic fluctuations, which are modeled as Ornstein--Uhlenbeck noises. For a single oscillator, the maximum output voltage and overall efficiency are attained for a soft piecewise-linear potential (providing a weak attractive constant force) but they are still fairly large for a harmonic potential. When several harmonic springs are linearly and bidirectionally coupled to form a ring, it is found that
counter-phase coupling can largely improve the performance while in-phase coupling worsens it.
Moreover, it turns out that few (two or three) coupled units perform better than more. }
\maketitle
\section{Introduction}
The objective towards downsizing and integration that began half a century ago with the invention of the transistor, necessarily implied developing suitable energy sources for the small, portable devices including wireless sensor systems, self-powered microelectronics, autonomous battery recharging, and many other applications. However, battery technology advances lagged behind compared to electronic technology. For instance, whereas disk storage density has increased over a thousand times in the last 35 years, the increase of battery energy density over that interval was only about three times \cite{Loreto}. Besides that, the major disadvantage of batteries is the need of replacing or recharging frequently once the charge is depleted. This could be a tedious and expensive procedure in the case of a large network consisting of hundreds of sensor nodes, and may be impossible if it is placed in a remote location.

For that reason, small energy-harvesting devices have been studied during the last few years as an alternative to batteries. These devices are capable of collecting the energy present in different forms in the environment. Naturally, the output of an energy harvester is not constant in time, and a power management circuit is needed before supplying the output to the circuits \cite{Vullers}. Moreover, in order to transform the harvested energy into electricity, the device must be equipped with a transducer.

There are many classifications for energy-harvesting devices, but the most usual ones focus on the energy source, e.g.\ kinetic, thermal, electromagnetic (light) \cite{Yildiz}. In \emph{kinetic} energy harvesting, the fundamental ingredient is the displacement or deformation of a moving part in the structure of the device. This motion can be converted into electric energy by suitable transduction mechanisms, usually electromagnetic (magnetic induction), electrostatic (capacity variation) or piezoelectric. In dealing with kinetic energy harvesting, \emph{piezoelectric} materials possess some advantages: larger power densities, high output voltage, simple structures, compatibility with micro-electro-mechanical systems (MEMS). A typical piezoelectric energy harvester consists of a cantilever, coupled to piezo-ceramic layers that generate alternating voltage output due to base excitation. A large number of models can be found in the literature, with their corresponding theoretical and experimental works \cite{Erturk,Sodano}.

As it initially applied to harvesting energy from ocean waves or machinery vibrations, most previous research has focused on the design of vibration resonators. Energy harvesters of this kind have peak performance when the excitation frequency matches their resonance frequency, but even a small detuning leads to substantial decrease of the output voltage. In most situations however, the vibrations in the environment are randomly distributed over a wide spectrum. Different approaches have been proposed to deal with this issue, as e.g.\ multimodal energy harvesting, resonance-tuning methods or frequency up-conversion \cite{Tang}. Several recent works \cite{ganv09,covg09,ganv10,litak1,litak2} have stressed \emph{non-linearity} as a resource to overcome this problem, specially if  the vibration statistics is broader than Gaussian \cite{dedw12,AFA,LAWNP}. Considering moreover that for the phenomenon of stochastic resonance \cite{rmp98,buga96}---a paradigm of the constructive interplay between stochasticity and nonlinearity---\emph{coupling} between units enhances the collective response \cite{Lindner1,Lindner2,Bouzat,Revelli,Izus,widl12}, here too one finds that coupling can boost the overall efficiency \cite{dedw13,55,56}.

This work continues those activities in studying the coupling effects, initiated in \cite{dedw13}. In Sec.\ \ref{sec:2}---after introducing the oscillator  model proposed as piezoelectric energy harvester (assumed to obey a bounded monostable potential of the form $U(x)=a_n\,|x|^n$ \cite{ganv10}) and the excitation force acting on it---the relationship between the system's performance and parameters $a_n$, $n$ is first explored. Next, the influence on the system's performance of linearly and bidirectionally coupling a set of harmonic oscillators is examined: the sign and strength of the coupling, and the number of coupled oscillators are optimized. Finally, the effect on the optimal configuration of further adjusting parameter $a_n$ (with the condition $a_n<1$) is analyzed. Section \ref{sec:3} collects our conclusions.

\section{Methods and results}\label{sec:2}
\subsection{Model}
In the spirit of \cite{ganv09,covg09,ganv10} (see also \cite{dedw12,AFA,LAWNP}) we start out by considering a one-dimensional inertial nonlinear oscillator $x(t)$---with mass $m$, damping constant $\gamma$, and governed by a monostable potential $U(x)$---coupled with strength $\sigma$ to a source of mechanical vibrations that produces an instantaneous force $\xi(t)$, and to a piezoelectric transducer. The latter provides a voltage $V(t)=K_c\,x(t)$, and reacts back on the oscillator with a force $K_vV(t)$. Constants $K_c$ (units of electric field) and $K_v$ (a linear charge density) are parameters of the piezoelectric device, which can be measured. The output voltage is in turn fed into a load circuit, with resistance $R$ and capacitance $C$, yielding a time constant $\tau_p=RC$.

The system is thus described by
\begin{eqnarray}
m\,\ddot{x}&=&-U'(x)-m\gamma\,\dot{x}-K_vV+\sigma\,\xi(t),\label{momento}\\
\dot{V}&=&K_c\,\dot{x}-V/\tau_p.
\end{eqnarray}

The source of mechanical vibrations $\xi(t)$---regarded as sto\-chastic but self-correlated or ``colored''---is modeled as an Ornstein--Uhlenbeck noise \cite{ganv09,covg09,ganv10,Mendez} with zero mean and self-correlation function
$$\langle\xi(t)\xi(t')\rangle=\tau^{-2}\exp[(t-t')/\tau].$$
As already stated, one of our goals is to infer the oscillator potential that maximizes efficiency. We restrict the search to the family considered in \cite{ganv10}, namely
\begin{equation}\label{potencial}
U(x)=a_n\,|x|^n,
\end{equation}
which become analytic for $n$ even. Here $a_n=U_0/|x_0|^n$, where $U_0$ has energy units and $x_0$ is a characteristic length, which can be taken as $\sqrt{\int\mathrm{d}x\,x^2\exp[-U(x)/\sigma^2]}$.

Being $V^2(t)/R$ the instantaneous power delivered to the load resistance, the measures of performance will be $V_\mathrm{rms}:=\langle V^2\rangle^{1/2}$---where $\langle V^2\rangle$ implies both a time-average during the observation interval and ensemble-average over noise realizations---and the efficiency (taken to be as defined in \cite{Mendez})
\begin{equation}
\eta=\eta_\mathrm{me}\eta_\mathrm{nm}=\frac1{R}\frac{\langle V^2\rangle}{\langle \dot{x}\,\xi\rangle},
\end{equation}
where $\eta_\mathrm{me}$ is the transducer's efficiency of converting mechanical to electrical power, and $\eta_\mathrm{nm}$ is the efficiency of power converted from the external noise to the power transferred from the oscillator to the transducer.


\subsection{Role of nonlinearity power n}
\begin{figure}
\vspace{-0.55cm}
\resizebox{1.1\columnwidth}{!}{\includegraphics{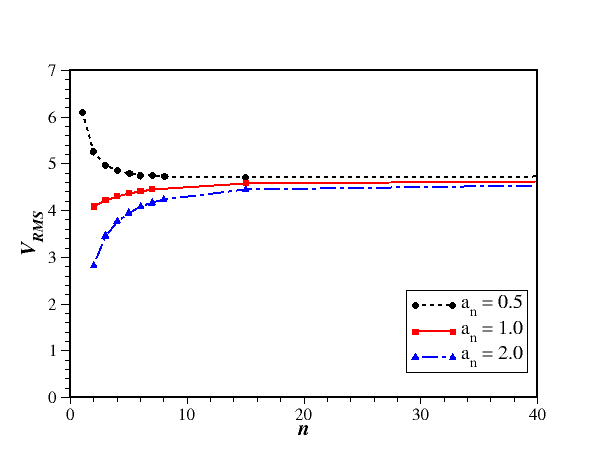}}
\resizebox{1.1\columnwidth}{!}{\includegraphics{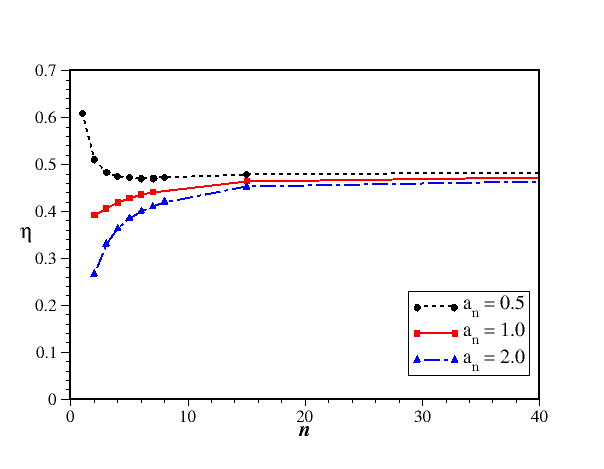}}
\caption{(Color online) Dependence of the mean output voltage $V_\mathrm{rms}$ (upper frame) and the overall efficiency $\eta$ (lower frame) on the potential's exponent $n$, for different amplitudes: $a_n=0.5$, 1 and 2. Fixed parameters: $m=1$, $\gamma= 1$,  $K_v= 1$, $K_c=1$, $\sigma =1$, $\tau=1$,  $\tau_p=2$.}
\label{fig1}
\end{figure}

Figure \ref{fig1} displays---for $a_n$ above, equal to and below one---$V_\mathrm{rms}$ (upper frame) and $\eta$ (lower frame) as functions of the exponent $n$ in Eq.\ (\ref{potencial}). The amplitudes of $a_n$ are kept constant throughout the exponents $n$ by considering the appropriate values of $U_0$ and $x_0$. With their own peculiarities, both performance indicators, $V_{rms}$  and $\eta$, follow the same trends:
\renewcommand{\theenumi}{\alph{enumi})\hspace{-.1cm}}
\begin{enumerate}
\item strong (very weak) dependence on $n$ for its lowest (high\-er) values,
\item performance improvement (worsening) at low $n$ for $a_n<1$ ($a_n\ge1$).
\end{enumerate}
In agreement with \cite{ganv10}, the best performance ($V_\mathrm{rms}=6.107$, $\eta=0.609$) is attained with the lowest $n$ and $a_n$ in this set, namely
\begin{equation}\label{n=1}
U(x)\propto\frac1{2}\,|x|.
\end{equation}
Note however that for $n=2$ (and $a_n=0.5$) both $V_\mathrm{rms}$ and $\eta$ are fairly high, possessing the additional advantage of the potential being analytic.

\subsection{Coupling several units}
The next step in optimizing the device is to find the most suitable linear coupling between units---as well as the optimal number of units---assuming periodic boundary conditions. It is our purpose here to thoroughly investigate preliminary evidence we have found that \emph{counter-phase} (anti-diffusive or anti-ferromagnetic like) coupling (i.e. coupling strength $k < 0$), outperforms in-phase one (i.e. $k > 0$, diffusive or ferromagnetic like), \cite{dedw13}.

\begin{figure}
\centering
\resizebox{1.1\columnwidth}{!}{\includegraphics{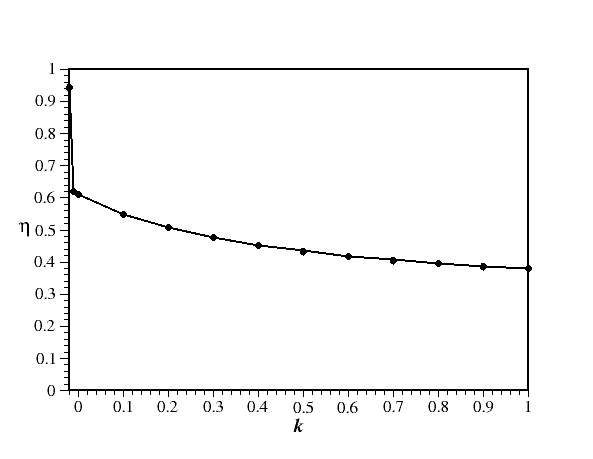}}
\caption{Efficiency $\eta$ as a function of the coupling strength $k$, for the piecewise-linear potential ($n = 1$). As $k$ becomes slightly negative, $\eta$ (and the mean output voltage $V_\mathrm{rms}$, not shown) \emph{diverge}. Fixed parameters: $m=1$, $\gamma= 1$,  $K_v= 1$, $K_c=1$, $\sigma =1$, $\tau=1$,  $\tau_p=2$.}
\label{lineal}
\end{figure}

The simplest generalization of Eq.\ (\ref{momento}) is
\begin{eqnarray}
m\,\ddot{x}_j&=&-U'(x_j)-m\gamma\,\dot{x}_j-K_vV_j+\sigma\,\xi_j(t)\nonumber\\
&&+k(x_{j+1}-2x_j+x_{j-1}),\label{acoplamiento}\\
\dot{V}_j&=&K_c\,\dot{x}_j-V_j/\tau_p,
\end{eqnarray}
$j=1,\ldots,N$. Now for the potential in Eq.\ (\ref{n=1}), the performance indicators (see e.g.\ Fig.\ \ref{lineal}) \emph{diverge} for some $k_\mathrm{di}<0$, with $|k_\mathrm{di}|\approx0.02$ \footnote{When $k=-0.01$, the efficiency $\eta$ scales up to 0.62 (Fig.\ \ref{lineal}), and $V_\mathrm{rms}$ reaches 6.23.}. That this divergence indicates the existence of a \emph{diffusive instability} \cite{xh,xg} is more clearly seen by writing Eq.\ (\ref{acoplamiento}) as
\begin{equation}
m\,\ddot{x}_j=-U'_\mathrm{eff}-m\gamma\,\dot{x}_j-K_vV_j+\sigma\,\xi_j(t)
\end{equation}
with
\begin{equation}
\small{U_\mathrm{eff}=\sum_{j}\left \{U(x_j)+\frac{k}{2}\left [(x_{j+1}-x_j)^2+(x_j-x_{j-1})^2\right ]\right \}},
\end{equation}
which has the continuous form
\begin{equation}
U_\mathrm{eff}=\int\mathrm{d}l\left\{U(x(l))+\frac{k}{2}(\partial_lx)^2\right\}.
\end{equation}
For $\sigma=0$, $k>0$ favors in-phase oscillation (uniform ground state) whereas $k<0$ favors counter-phase oscillation (finite-wavevector ground state). If by effect of noise, a given oscillator performs a large excursion (so producing a large piezoelectric voltage), $k>0$ (diffusive coupling) will tend to smooth it up, whereas $k<0$ (antidiffusive coupling) will tend to enhance it.

In order to perform a deeper analysis with the piece\-wise-linear potential oscillators of Eq.\ (\ref{n=1}) and Fig.\ \ref{lineal}, high\-er-order corrections (that is, including terms beyond the diffusive one) should be introduced to the definition of the coupling in Eq.\ (\ref{acoplamiento}) \cite{xh,xg,wio}. Before embarking on such a task however, it is worth exploring the dependence of $k_\mathrm{di}$ on $n$ given that (according to Fig.\ \ref{fig1}) the uncoupled performance is larger for the lowest $n$ values than for the rest, being  still very good in the harmonic case $n=2$.

\begin{figure}
\resizebox{1.1\columnwidth}{!}{\includegraphics{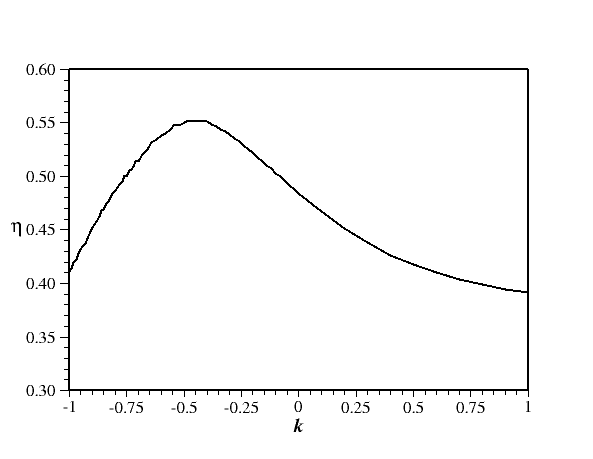}}
\caption{For $n>2$, no divergence shows up in either $\eta$ or $V_\mathrm{rms}$ for $k<0$. Fixed parameters: $m=1$, $\gamma= 1$,  $K_v= 1$, $K_c=1$, $\sigma =1$, $\tau=1$,  $\tau_p=2$.}
\label{n=3}
\end{figure}

In the coupled case, no diffusive instability shows up for $n>2$. Already for $n=3$, a maximum is observed in the performance indicators at some $k_\mathrm{max}\approx-0.5$, with $\eta=0.552$ (Fig.\ \ref{n=3}) and $V_\mathrm{rms}=5.492$ (not shown) \footnote{For $n=4$, $\eta=0.508$ and $V_\mathrm{rms}=5.074$.}. There is still a diffusive instability for $n=2$ (see Fig.\ \ref{n=2} below, for the case of three coupled harmonic oscillators), but $k_\mathrm{di}$ is shifted toward safely larger values, allowing the efficiency $\eta$ (Fig.\ \ref{n=2}) and the mean output voltage $V_\mathrm{rms}$ (not shown) to largely exceed the maximum values attained for $n>2$.

\subsection{Number of units}
After confirming our preliminary evidence \cite{dedw13} that the performance of  \emph{antidiffusively} coupled units can be notably enhanced as compared with the non-coupled case, we ask ourselves what the optimal number $N$ of (antidiffusively) coupled units is. For our surprise however (see Fig.\ \ref{numberofunits}), the best performance is attained when $N$ is very low (two or three). Afterwards, a kind of plateau is reached. Configurations whose number of units is \emph{odd} perform slightly better than those for which it is even, because they yield on average a net displacement. As expected, the value of $k_\mathrm{max}$ depends on $N$.

\begin{figure}[!h]
\resizebox{1.1\columnwidth}{!}{\includegraphics{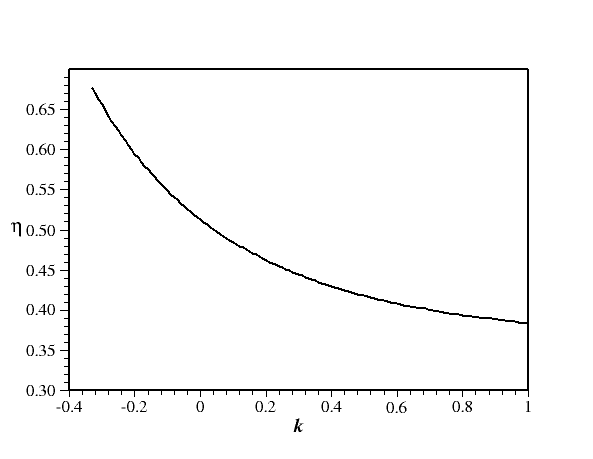}}
\caption{Three coupled harmonic oscillators. The divergence shows up at larger $|k|$ and for $k=-0.33$, the system attains $\eta= 0.677$ and $V_\mathrm{rms}= 6.778$. Fixed parameters: $m=1$, $\gamma= 1$,  $K_v= 1$, $K_c=1$, $\sigma =1$, $\tau=1$,  $\tau_p=2$.}
\label{n=2}
\end{figure}

\begin{figure}
\resizebox{1.1\columnwidth}{!}{\includegraphics{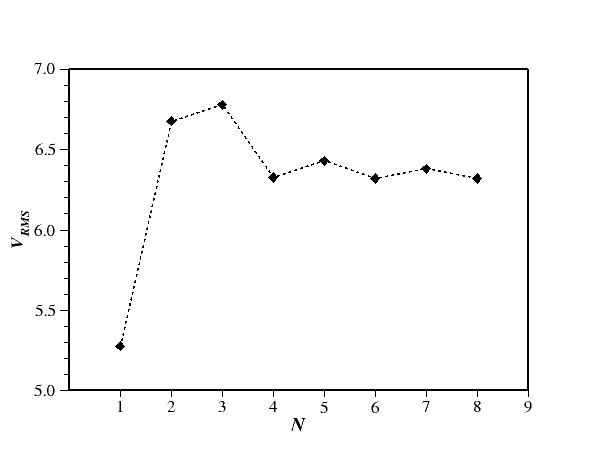}}
\resizebox{1.1\columnwidth}{!}{\includegraphics{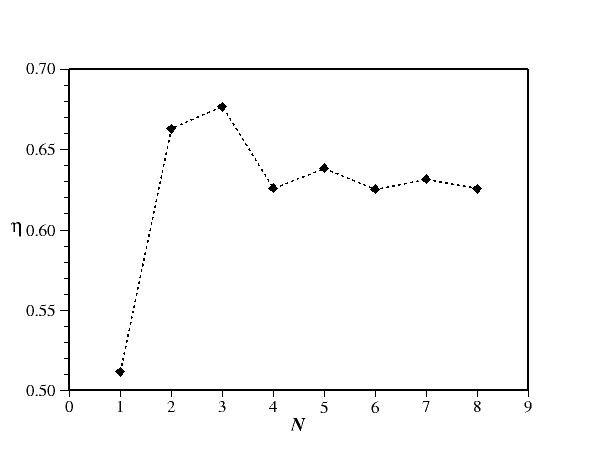}}
\caption{Optimal performance of coupled harmonic oscillators. Upper frame: mean output voltage $V_\mathrm{rms}$; lower frame: overall efficiency $\eta$. Fixed parameters: $m=1$, $\gamma= 1$,  $K_v= 1$, $K_c=1$, $\sigma =1$, $\tau=1$,  $\tau_p=2$.}
\label{numberofunits}
\end{figure}


\subsection{The role of the potential softness}

Once the optimal configuration ($N=3$, $a_n<1$) has been found for $n=2$, we decrease $a_n$ further in order to check whether the performance keeps improving, and find the minimum value of $a_n$ for which the negative couplings defined by Eq.\ (\ref{acoplamiento})---i.e.\ without considering higher order corrections---lose meaning.

\begin{figure}
\hspace{1.cm}
\resizebox{0.9\columnwidth}{!}{\includegraphics[trim=50 0 0 0,width=0.55\textwidth]{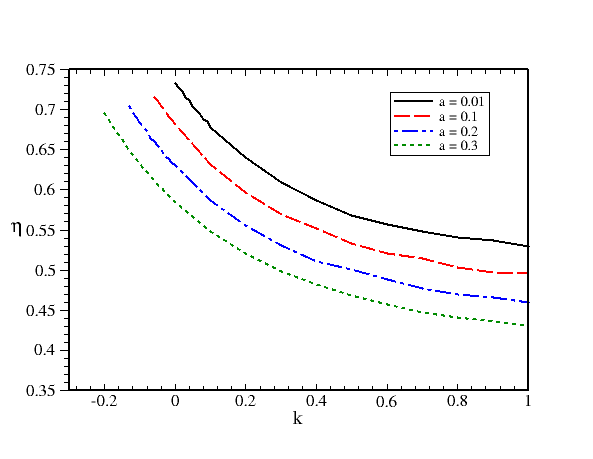}}
\caption{(Color online) Efficiency $\eta$ as a function of coupling strength $k$ for different values of potential amplitudes $a_n$. Fixed parameters: $m=1$, $\gamma=1$,  $K_v=1$, $K_c=1$, $\sigma =1$, $\tau=1$,  $\tau_p=2$.}
\label{n=8}
\end{figure}

We have found that $|k_\mathrm{di}|$ decreases as $a_n$ does. The limit seems to be around $a_n=0.01$, beyond which every negative coupling between units leads to a divergence in $\eta$ and $V_\mathrm{rms}$ (clearly, a small $a_n$  implies either a small $U_0$  or a large $x_0$).


\section{Discussion}\label{sec:3}

For a model oscillator proposed as an energy harvester through piezoelectric conversion, we have analyzed monostable potentials of the form $U(x)=a_n\,|x|^n$. In agreement with \cite{ganv10}, the best performance is attained with $n=1$ and $a_n<1$; namely a soft piecewise-linear potential, providing a weak attractive constant force. The performance indicators strongly decay with $n$ for its few lowest values, but they are still high enough for the harmonic oscillator.

We have studied the effect of coupling between units and sought the optimal configuration (number $N$ of coupled units) in order to enhance the system's energy harvesting. In agreement with \cite{dedw13}, whilst diffusive coupling between units reduces the system's performance, anti-diff\-usive
couplings cause an enhancement. Next we found that it does so via the mechanism of a \emph{diffusive instability}. As a metaphor, we can picture this situation as an anti-ferromagnetic coupling between spins \cite{am}, or an inhibitory coupling in neuron systems \cite{da}.

Nevertheless, the main point to be made is how to implement such a form of coupling. The foregoing results are just indicative, and should be regarded as a ``toy model''. However, even such a simple setup depicts the fact that corrections going beyond a diffusive coupling need to be added in order to globally control the instability \cite{xh,xg,wio}. This will be subject of further studies, together with more elaborate aspects and/or models.
\vskip.5cm

Support by the following institutions is acknowledged: CSIC (Spain) under i-COOP+ program, MINECO (Spain), under project No. FIS2014-59462-P, by HSW; CONICET and UNMdP (Argentina), by JIPR and RRD.

\bibliographystyle{epj}
\bibliography{coupledHO}

\begin{thebibliography}{31}

\bibitem{Loreto}
L.~Mateu, F.~Moll, \emph{Review of energy harvesting techniques and
  applications for microelectronics (keynote address)}, in \emph{Proc. SPIE
  5837, VLSI Circuits and Systems II} (2005), p. 359

\bibitem{Vullers}
R.J.M. Vullers, R.~{van Schaijk}, I.~Doms, C.~{Van Hoof}, R.~Mertens, Solid-St.
  Electron. \textbf{53}, 684 (2009)

\bibitem{Yildiz}
F.~Yildiz, J. Technol. Studies \textbf{35}, 40 (2009)

\bibitem{Erturk}
A.~Erturk, Ph.D. thesis, Virginia Polytechnic Institute and State University
  (2009)

\bibitem{Sodano}
H.A. Sodano, D.J. Inman, G.~Park, J. Intell. Mater. Syst. Struct. \textbf{16},
  799 (2005)

\bibitem{Tang}
L.~Tang, Y.~Yang, C.K. Soh, J. Intell. Mater. Syst. Struct. \textbf{21}, 1867
  (2010)

\bibitem{ganv09}
L.~Gammaitoni, I.~Neri, H.~Vocca, Appl. Phys. Lett. \textbf{94}, 164102 (2009)

\bibitem{covg09}
F.~Cottone, H.~Vocca, L.~Gammaitoni, Phys. Rev. Lett. \textbf{102}, 080601
  (2009)

\bibitem{ganv10}
L.~Gammaitoni, I.~Neri, H.~Vocca, Chem. Phys. Lett. \textbf{375}, 435 (2010)

\bibitem{litak1}
G.~Litak, E.~Manoach, Eur. Phys. J. Special Topics \textbf{{\textbf 222}}, 1479
  (2013)

\bibitem{litak2}
G.~Litak, E.~Manoach, E.~Halvorsen, Eur. Phys. J. Special Topics
  \textbf{{\textbf 224}}, 671 (2015)

\bibitem{dedw12}
J.I. Deza, R.R. Deza, H.S. Wio, Europhys. Lett. \textbf{100}, 38001 (2012)

\bibitem{AFA}
J.I. {Pe\~na Rossell\'o}, J.I. Deza, H.S. Wio, R.R. Deza, Anales AFA
  \textbf{25}, 54 (2014)

\bibitem{LAWNP}
J.I. {Pe\~na Rossell\'o}, R.R. Deza, J.I. Deza, H.S. Wio, Papers in Physics
  \textbf{7}, 070014 (2015)

\bibitem{rmp98}
L.~Gammaitoni, P.~H{\"a}nggi, P.~Jung, F.~Marchesoni, Rev. Mod. Phys.
  \textbf{70}, 223 (1998)

\bibitem{buga96}
A.R. Bulsara, L.~Gammaitoni, Phys. Today \textbf{49}, 39 (1996)

\bibitem{Lindner1}
J.F. Lindner, B.K. Meadows, W.L. Ditto, M.E. Inchiosa, A.R. Bulsara, Phys. Rev.
  Lett. \textbf{75}, 3 (1995)

\bibitem{Lindner2}
J.F. Lindner, B.K. Meadows, W.L. Ditto, M.E. Inchiosa, A.R. Bulsara, Phys. Rev.
  E \textbf{53}, 2081 (1996)

\bibitem{Bouzat}
H.S. Wio, S.~Bouzat, B.~{von Haeften}, Physica A \textbf{306}, 140 (2002)

\bibitem{Revelli}
H.S. Wio, J.A. Revelli, M.A. Rodr\'{\i}guez, R.R. Deza, G.G. Iz\'us, Eur. Phys.
  J. B \textbf{69}, 71 (2009)

\bibitem{Izus}
B.~{von Haeften}, G.G. Iz\'us, H.S. Wio, Phys. Rev. E \textbf{72}, 021101
  (2005)

\bibitem{widl12}
H.S. Wio, R.R. Deza, J.M. L\'opez, \emph{An Introduction to Stochastic
  Processes and Nonequilibrium Statistical Physics, \emph{revised edition}}
  (World Scientific, Singapore, 2012)

\bibitem{dedw13}
J.I. Deza, R.R. Deza, H.S. Wio, Nanoenergy Lett. \textbf{6}, 29 (2013)

\bibitem{55}
P.~H{\"a}nggi, M.~Inchiosa, D.~Fogliatti, A.~Bulsara, Phys. Rev. E \textbf{62},
  6155 (2000)

\bibitem{56}
J.~Casado-Pascual, J.~{G\'omez-Ordo\~nez}, M.~Morillo, P.~H{\"a}nggi, Phys.
  Rev. E \textbf{67}, 036109 (2003)

\bibitem{Mendez}
V.~M\'endez, D.~Campos, W.~Horsthemke, Phys. Rev. E \textbf{88}, 022124 (2013)

\bibitem{xh}
M.C. Cross, P.C. Hohenberg, Rev. Mod. Phys. \textbf{65}, 851 (1993)

\bibitem{xg}
M.~Cross, H.~Greenside, \emph{Pattern formation and dynamics in nonequilibrium
  systems} (CUP, Cambridge, 2009)

\bibitem{wio}
H.S. Wio, Int. J. Bif. Chaos \textbf{19}, 2813 (2009)

\bibitem{am}
N.W. Ashcroft, D.N. Mermin, \emph{Solid state physics} (Harcourt, Orlando, FL,
  1976)

\bibitem{da}
P.~Dayan, L.F. Abbott, \emph{Theoretical neuroscience} (MIT Press, Cambridge,
  MA, 2001)

\end{thebibliography}
\end{document}